\documentclass{llncs}
\usepackage[utf8]{inputenc}
\usepackage{graphicx}
\usepackage{caption}
\begin{document} 

\title{Low Cost High Integrity Platform \\\small{regular paper}}
\titlerunning{Formal Method in Industry}  
%
\author{Thierry Lecomte\inst{1} \and 
David Deharbe\inst{1} \and 
Denis Sabatier\inst{1} \and 
Etienne Prun\inst{1} \and 
Patrick Péronne\inst{1} \and
Emmanuel Chailloux\inst{2} \and
Steven Varoumas\inst{2} \and
Adilla Susungi\inst{2} \and
Sylvain Conchon\inst{3}
}
\authorrunning{Thierry Lecomte et al.} 
%
\tocauthor{Thierry Lecomte, David Deharbe, Denis Sabatier, Etienne Prun, and Patrick Peronne}
\institute{
CLEARSY Systems Engineering, Aix en Provence, France\\
\email{thierry.lecomte@clearsy.com} 
\and
Sorbonne Université, CNRS, LIP6, Paris, France\\
\email{emmanuel.chailloux@lip6.fr}
\and
Université Paris-Sud, LRI, Orsay, France\\
\email{sylvain.conchon@lri.fr} 
}

\maketitle              

\appendix 
\section{Revolution for developing of safety critical application}
Developing safety critical applications often requires rare human resources to complete successfully while off-the-shelf block solutions appear difficult to adapt especially during short-term projects. 
Developed during the R\&D project FUI LCHIP\cite{Lecomte01}, the CLEARSY Safety Platform fulfills a need for a technical solution to overcome the difficulties to develop SIL3/SIL4 system. Its technology is based on a smart combination of diverse hardware (2x PIC 32 micro-controllers) and a formal method with proof heavily used in the railways industry for decades.
It avoids most testing and ensures safety at the highest level.  

The CLEARSY Safety Platform is both a software and a hardware platform aimed at designing and executing safety critical applications. 
One formal modelling language (B) is used to program the board. Programs are developed using a dedicated IDE or could be the by-product of some translation from a Domain Specific Language to B. 
The IDE takes care of the verification of the software (type check, proof, compilation) and then ensures its uploading to the hardware platform. Program is guaranteed to execute until a misbehaviour is detected, leading to a safe restricted mode where board outputs are deactivated.\\

\paragraph{Added value}The CLEARSY Safety Platform eases the development of safety critical applications as:
\begin{itemize}
    \item it covers the whole development cycle of control-command systems based on digital inputs/outputs.
    \item development time is shortened as the safety principles are built-in, and are out of reach of the developer who cannot alter them. Development is focused on the behaviour.
    \item the testing phase is dramatically reduced as the mathematical proof replaces unit and integration testing (based on a formal language (B) and related proof tools).
\end{itemize}

\paragraph{Eased certification}The CLEARSY Safety Platform eases the certification of safety critical applications as the safety cannot be altered by the developer. It comes with a certification kit to be used for the safety case of the system embedding the CLEARSY Safety Platform. The building blocks of the CLEARSY Safety Platform, already certified in international railway projects (platform screen doors controller in Brazil (São Paulo line 15) and Sweden (Stockholm Citybanan), remote IOs in Canada), have been used to develop a generic version of this technology that could fit a broader range of applications.

\section{Technology}
\begin{figure}[h]
\centering\includegraphics[scale=0.20]{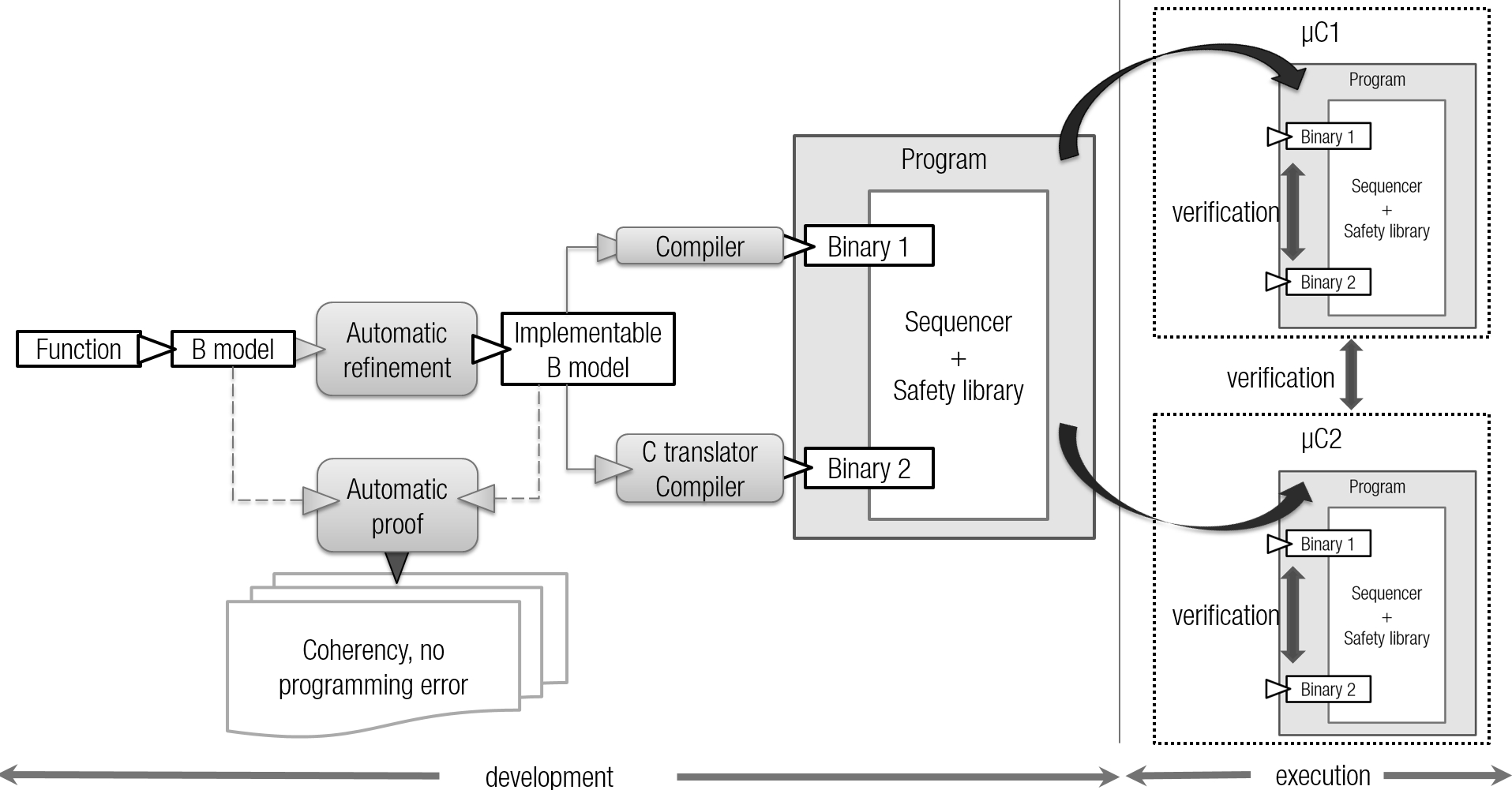}
\caption{Full path from function description to safe execution with the CLEARSY Safety Platform. Round boxes are tools, rectangular boxes are files.}
\label{arch:principes}
\end{figure}

With the CLEARSY Safety Platform, the very technical aspects related to safety are taken into account by the platform, leaving the developer to focus only on the development of the function to perform. The CLEARSY Safety Platform is made of two parts: an IDE to develop the software and an electronic board to execute this software.  The full process is described in figure \ref{arch:principes}.

\paragraph{Software development}It starts with the function specification (natural language) to develop. The developer has to provide a B model of it (specification and implementation) using the schema:
\begin{itemize}
    \item the function to program is a loop, where the following steps are performed repeatedly in sequence:
\begin{itemize}
    \item the inputs are read.
    \item some computation is performed.
    \item the outputs are set.
\end{itemize}
\item The steps related to inputs and outputs are fixed and cannot be modified. 
\item Only the computation may be modified to obtain the desired behaviour.
\end{itemize}
The implementation is usually handwritten but could also be generated automatically with the B Automatic Refinement Tool. The B models are proved (mostly automatically as the level of abstraction of typical command \& control applications is low) to be coherent and to contain no programming error.
From the implementable model, two binaries are generated:
\begin{itemize}
    \item binary$_1$, obtained via a dedicated compiler, developed by CLEARSY, transforming a B model into HEX file,
    \item binary$_2$, produced with the Atelier B C code generator then compiled with the GCC compiler into another HEX file.
\end{itemize} 
Each binary represents the same function but is supposed to be made of different sequences of instructions because of the diversity of the tool chains. Then the two binaries binary$_1$ and binary$_2$ are linked with:
\begin{itemize}
    \item a sequencer, in charge of reading inputs, executing binary$_1$ then binary$_2$, then setting the outputs
    \item a safety library, in charge of performing safety verification (more details in https://www.clearsy.com/en/download/download-documentation). In case of failing verification, the board enters panic mode, meaning the outputs are deactivated (no power is provided to the Normally Open (NO) outputs, so the output electric circuits are open), the board status LED start flashing, and the board enters an infinite loop doing nothing. A hard reset (power off or reset button) is the only possibility to interrupt this panic mode.
\end{itemize}
The final program is thus made of binary$_1$, binary$_2$, the sequencer and the safety library. The memory mappings of binary$_1$ and binary$_2$ are separate. This program is then uploaded on the two micro-controllers $\mu C_1$ and $\mu C_2$. 

\paragraph{Verification}The bootloader, on the electronic board, checks the integrity of the program (CRC, separate memory spaces). Then both micro-controllers start to execute the program. During its execution, the following are performed:
\begin{itemize}
    \item internal verification:
\begin{itemize}
    \item every cycle, binary$_1$ and binary$_2$ memory spaces (variables) are compared;
    \item regularly, binary$_1$ and binary$_2$ memory spaces (program) are compared in deferred mode;
    \item regularly, the identity between memory output states and physical output states is checked to detect if the board is unable to command the outputs.
\end{itemize} 
\item external verification:
\begin{itemize}
    \item regularly (every 50ms at the latest), memory spaces (variables) are compared between $\mu C_1$ and $\mu C_2$. 
\end{itemize} 
\end{itemize} 
If any of these verifications fail, the board enters the panic mode.

\begin{figure}[h]
\centering\includegraphics[scale=0.35]{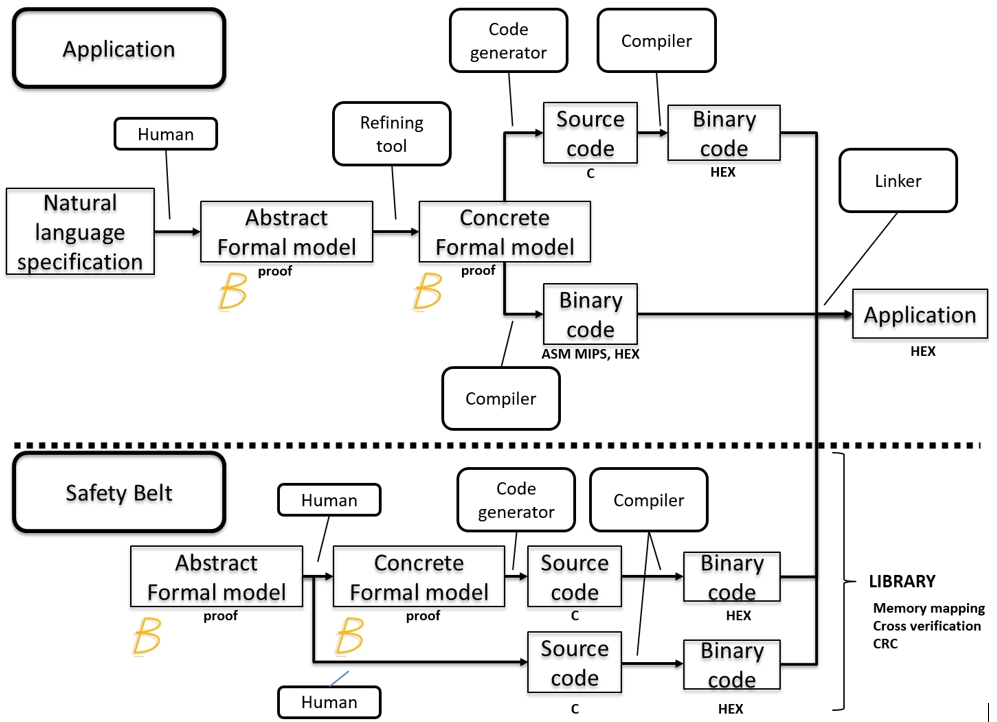}
\caption{Tools and files involved in the generation of the software}
\label{arch:process}
\end{figure}

\paragraph{Tools}The whole process is fully supported by adequate tools. In the figure \ref{arch:process}, the tools and text/binary files generated are made explicit for both the application (path used every time an application is developed) and the safety belt (developed once for all by the IDE development team. Note that from the abstract formal model, one part of the software is developed in B with concrete formal model, while the other part is developed manually. It happens when using B provides no added-value (for example low-level IO). A component modelled in B and implemented manually is called a basic machine. \\
The tools are issued from Atelier B, except:
\begin{itemize}
    \item the B to HEX compiler, initially developed to control platform screen doors for metro lines in Brazil. This tool proceeds in two steps: a translation from B to ASM MIPS, then from ASM MIPS to HEX. 
    \item the C to HEX GCC compiler.
    \item the linker combining the 2 hex files with the safety sequencer and libraries.
    \item the bootloader.
\end{itemize}

\paragraph{Safety principles}The safety is built on top of few principles:
\begin{itemize}
    \item a B formal model of the function to develop, proved to be coherent, to correctly implement its specification, and to be programming error-free,
    \item four instances of the same function running on two micro-controllers (two per micro-controller with different binaries obtained from diverse tool-chains) and the detection of any divergent behaviour among the four instances,
    \item the deferred cross-verification of the programs on the two $\mu C$,
    \item outputs require both $\mu C_1$ and $\mu C_2$ to be alive and running as one provides energy and the other one the command,
    \item output physical states are regularly verified to comply with the memory states, to check the ability of the board to command its outputs,
    \item input signals are continuous (0 or 5V) and are made dynamic (addition of a frequency signal) in order to prevent the short-circuit current from being considered  as high level (permissive) logic.
\end{itemize}

From a safety point of view, the current architecture is valid for any kind of mono-core processor. The decision of using PIC32 micro-controllers (able to deliver around 50 DMIPS) was made based on our knowledge and experience of this processor. Implementing the CLEARSY Safety Platform on other hardware 
would "only" require the existing electronic board and software tools to be modified, without impacting much the safety demonstration.
 
\section{Complementary technologies}

\begin{figure}[h]
\centering\includegraphics[scale=0.26]{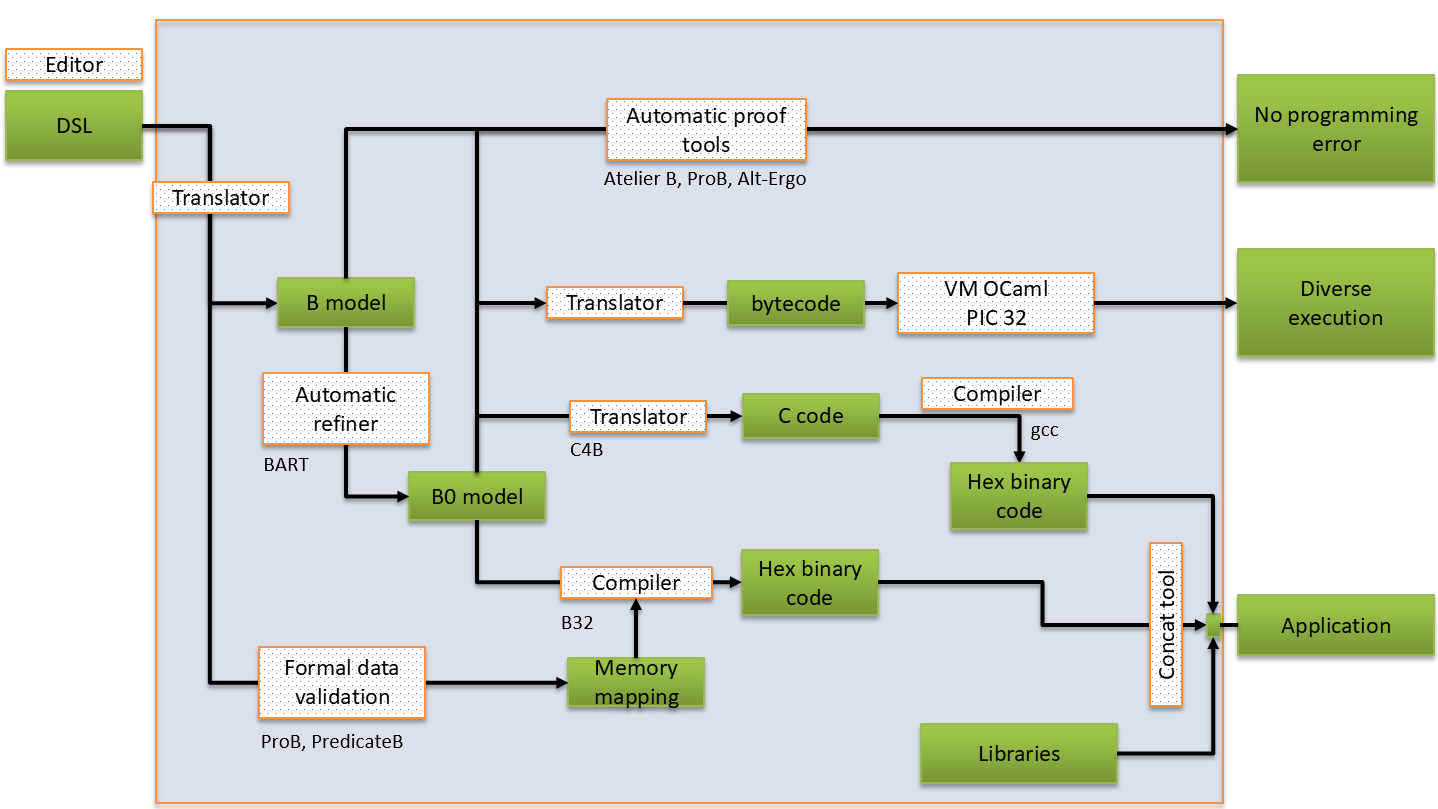}
\caption{Tools architecture and dependencies}
\label{arch}
\end{figure}

The CLEARSY Safety Platform implements a well-oiled automatic process from a proved B model to a safe execution. However several limitations prevent a larger exploitation:
\begin{itemize}
    \item the B language (even restricted to a subset) is not widely disseminated and often considered as difficult to use by engineers.
    \item the fully automatic model proof is only achieved with "bounded complexity algorithms", even if we are not considering implementing  metro automatic pilots with the CLEARSY Safety Platform.
    \item fine-tuning and debugging applications that run simultaneously on two processors are difficult to achieve.
\end{itemize}
To overcome these limitations, several features (see figure \ref{arch} for the global picture) have been added to the CLEARSY Safety Platform and are detailed in the forthcoming sections:
\begin{itemize}
    \item connection with Domain Specific Languages, to entitle engineers to model with their usual language and to keep the formalities behind the curtain. 
    \item improved proof performances. That way, the formal verification is also hidden behind the curtain, as log as the complexity of the implemented algorithms is compatible with the automatic proof capabilities (we are not aimed at metro automatic pilots here)
    \item debugging facilities on host with a dedicated VM
\end{itemize}

\subsection{Connection with DSL}
\begin{figure}[h]
\centering\includegraphics[scale=0.24]{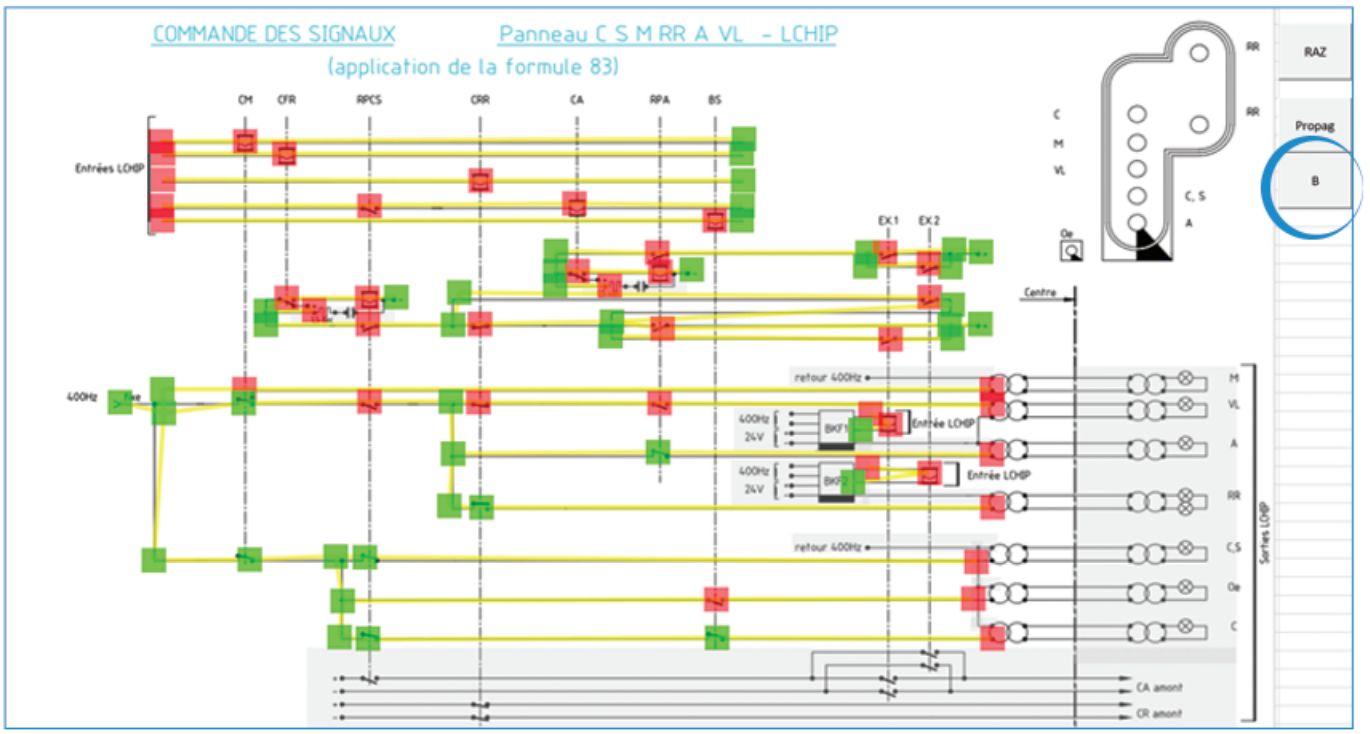}
\caption{On-the-fly automatic model extraction from relay-based schematics and B model generation}
\label{rssr}
\end{figure}

A B model is the mandatory entry-point of the CLEARSY Safety Platform. However this B model could be 
obtained from the translation of a DSL model into B. This approach allows to seamlessly involve domain experts without changing their modelling languages.
Experiments were conducted with SNCF \cite{Deharbe19} in order to translate relay-schemes into B models for various applications (ITCS - wrong track temporary installation, signal controller). The translation allows to interactively describes relay-schemes and to generate formal models (figure \ref{rssr}).

Finally a wired-logic installed for decades could be easily replaced by a safe programmed-logic. The poster "Porting Relay-Based Schemas to a SIL4 Programmable Control Platform" is available at https://www.clearsy.com/wp-content/uploads/2019/07/Affiche-A1-v3.pdf.

Other DSLs are being considered, such as grafcet / sequential functional / chart to directly address automation outside railways.

\subsection{Improved Proof}

The B method enforces that the development is mathematically sound by producing proof obligations. A proof obligation has
a goal and hypotheses, all expressed in a mathematical formalism (first order logic, with set theory and integer arithmetic).
Historically, proof obligations have been verified using a mix of automatic proof procedures provided in Atelier B and user 
insight (e.g. case splitting, quantifier instantiation, etc.) All proof related tools are part of the IDE and have been in
place since the inception of Atelier B. 

In order to benefit from the technical and scientific improvements in the field of automated reasoning, a new proof obligation
generator has been developed. Proof obligations are then produced in a XML-based format which can then be translated to the
native format of any formal verification tool. As a proof-of-concept, a connection to third-party provers was conducted as part
of the B-Ware initiative, using the Why3 platform~\cite{boogie11why3} as a gateway.
This experiment produced
excellent results in terms of proof automation, in particular for the Alt-Ergo prover~\cite{Conchon2014}. The results of this
work are now being integrated in Atelier B to provide new proof features to the IDE. Using this approach, we target full proof 
automation for selected DSLs.

Cubicle \cite{ConchonGKMZ12} is a model checker for verifying safety properties of transition systems manipulating arrays. Cubicle finds inductive invariants of systems by the integration of the SMT solver Alt-Ergo with a backward reachability algorithm. In order to improve its reachability algorithm, we developed a new approach based on program unwinding, reminiscent to property-driven reachability.
This algorithm has then been extended for reasoning about weak memory model for verifying properties of assembly x86-TSO programs.
Our extension relies on a tuned axiomatic memory model for SMT and a specific backward reachability algorithm that exploits a new partial order reduction technique for the TSO model of x86. First experiments benchmarks from synchronization barriers to mutual exclusion implementation like the spinlock from Linux 2.6 kernel show that this approach is very promising.

\subsection{A Virtual Machine Approach}

One original contribution of the LCHIP project is the possibility to run the embedded program in a virtual machine (VM) of a high-level programming language. This primarily provides an alternative mean of execution, which offer new ways of debugging and improving security. Indeed, the original compilation models considered (which target the C language or the PIC assembly) follow similar patterns, but with the use of a VM we can produce binaries with a very different execution path, ensuring redudancy checks of the behaviour of programs. To this end, we thus propose a specific implantation of the virtual machine of the OCaml language. This machine takes advantages from the expressiveness of this language (implementing functional, imperative, object-oriented and modular paradigms), as well as its guarantees (such as the safety provided by its static type system).

A first experience of implementing this VM for PIC18, called OCaPIC \cite{PADL2015}, has shown the feasibility of running complex  programs on micro-controllers with scarce resources (a few KiB for RAM, 32KiB for flash memory). This approach was then generalized with the OMicroB \cite{ERTS2018} system, which starts from a byte-code executable program for the OCaml VM and embeds the byte-code in a C file. This generic approach provides the possibility to compile and load a given program on different families of micro-controllers (AVR, ARM, PIC32, \dots). To do this, several types of optimization are necessary both to reduce the size of the byte-code  and for automatic memory management. OMicroB produces an binary that embeds the byte-code interpreter, the execution library and the byte-code. Note that since OMicroB compilation path uses the C language, like the previously described compiling path that consists of translating B to C, security concerns would entail the use of another C compiler in order to dissipate probable unknown bugs inherent to the chosen C compiler.

The OMicroB system has successfully been ported for the LCHIP hardware, and various OCaml programs have been executed on the platform. Debugging can be rendered easier by using OMicroB since it comes with a simulator that can represent interactions with the external hardware to which is connected the microcontroller(s). Furthermore, for possible modifications of the hardware (as discussed in section B), this approach is easily adaptable, since it has been designed to be portable on many architectures. 

In order to run the program derived from the B model, we need to provide a source-to-source translator from B0 to OCaml, as it is shown in figure \ref{arch}. This translator will be very close to the C4B translator that produces a C source code, since we can take advantage of all the imperative constructs of the OCaml language (by using mutable records and arrays as equivalent to C variables and data structures, for example).

As other future works, the internal checks between the two binaries could be orchestrated by a synchronous extension "à la Lustre" called OCaLustre \cite{ERTS2016}. Since this extension is built above the OCaml language, it is completely compatible with our virtual machine. An additional interest of this virtual machine approach is to factorize in property checks at the byte-code level, such as the estimation of the WCET (Worst Case Execution Time) for programs that do not dynamically allocate memory (like synchronous programs) \cite{WCET2019}. This is currently possible on hardware where the memory model is simple (no caches, simple scalar pipelines), like some AVR microcontrollers, and thus guarantees \emph{timing compositionality} of each (bytecode) instruction. On simple hardware, we can herewith project the byte-code analysis to the actual architecture instructions set. However, the application of such a computation on more complex hardware like the PIC32 (that has pre-fetch cache system) is still under consideration.

\section{Education, dissemination and exploitation}

The CLEARSY Safety Platform is being used in several universities in Europe and America, to teach both formal methods and embedded systems \footnote{More information at https://www.clearsy.com/en/category/clearsy-safety-platform-en/}. It is an interesting mean to involve mathematicians into close-to-hardware thematic, and conversely to bring embedded systems practitioners to more abstract reasoning. Lectures were given up to Master2, aimed at demonstrating safety engineering at computer scientists and electronic engineers, with dedicated hands-on sessions.

\begin{figure}[h]
\centering\includegraphics[scale=0.06]{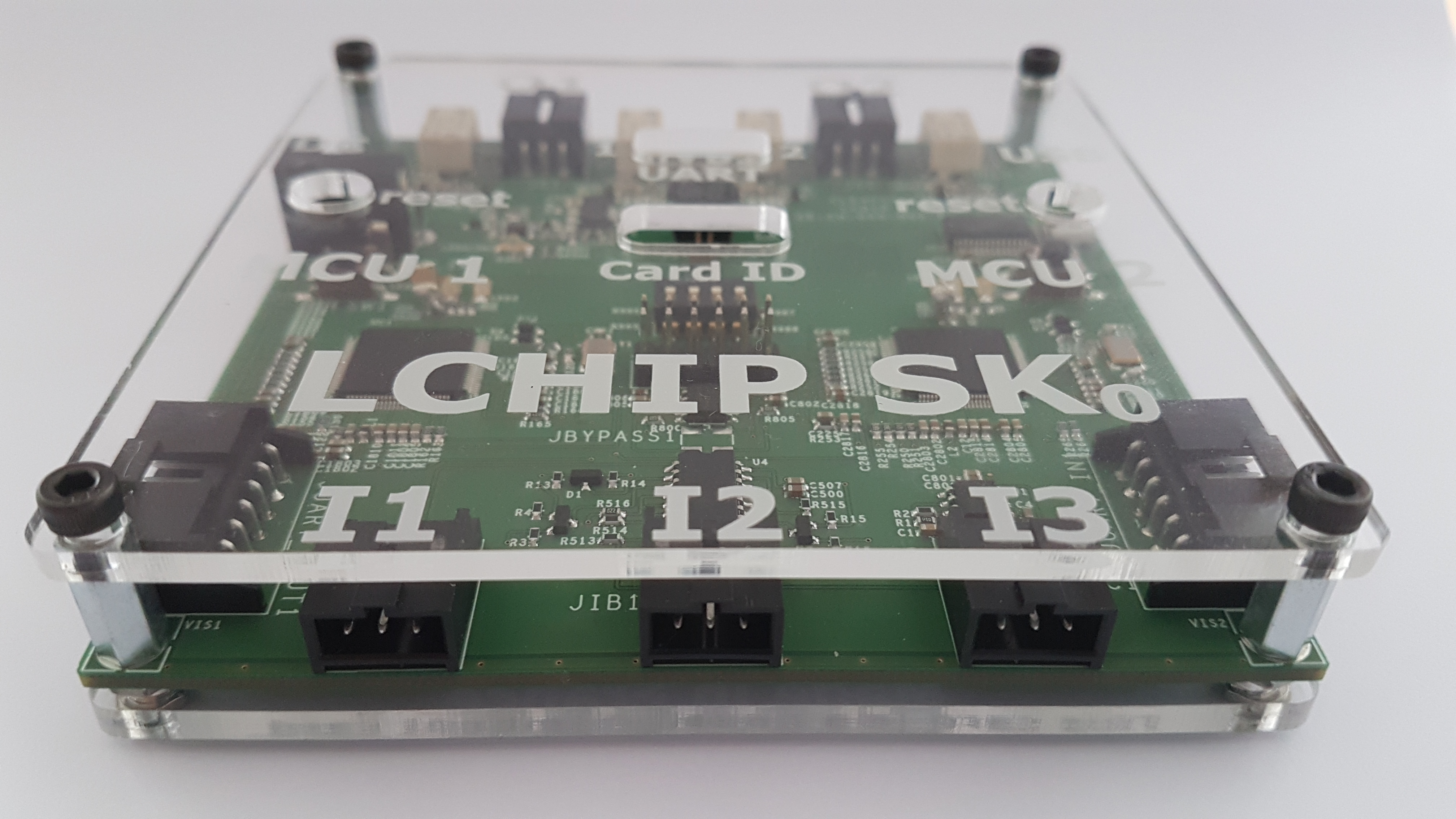}
\caption{The CLEARSY Safety Platform Starter Kit 0}
\label{SK0}
\end{figure}

The two existing starter kits SK$_0$ and SK$_1$ are clearly aimed at respectively education and prototyping. A forthcoming industry-strength version will be provided as a daughter-board (PLC without inputs/outputs) to integrate into in-house development.
Easier certification, lower development and deployment costs will have a dramatic impact on public safety, allowing to embed safety in systems for a limited cost. 
More over on-going research projects are aimed at bringing safety to robotic systems. 

\bibliographystyle{splncs03}
\bibliography{biblio}

\begin{thebibliography}{1}
\providecommand{\url}[1]{\texttt{#1}}
\providecommand{\urlprefix}{URL }

\bibitem{boogie11why3}
Bobot, F., Filli\^atre, J.C., March\'e, C., Paskevich, A.: Why3: Shepherd your
  herd of provers. In: Boogie 2011: 1st Int'l Workshop on Intermediate
  Verification Languages. pp. 53--64. Wroc\l{}aw, Poland (August 2011)

\bibitem{ConchonGKMZ12}
Conchon, S., Goel, A., Krstic, S., Mebsout, A., Zaïdi, F.: Cubicle: {A}
  parallel {SMT}-based model checker for parameterized systems - tool paper.
  In: Computer Aided Verification - 24th International Conference, {CAV} 2012,
  Berkeley, CA, USA, July 7-13, 2012 Proceedings. pp. 718--724 (2012)

\bibitem{Conchon2014}
Conchon, S., Iguernelala, M.: Tuning the Alt-Ergo SMT Solver for B Proof
  Obligations, pp. 294--297. Springer (2014)

\bibitem{Deharbe19}
Dalay~Pereira, David~Deharbe, M.P.P.B.: B-specification of relay-based railway
  interlocking systems based on the propositional logic of the system state
  evolution. In: Reliability, Safety and Security of Railway Systems, {IJCAR}
  2019, Lille, France, June 4 - 6, 2019, Proceedings (2016)

\bibitem{Lecomte01}
Lecomte, T.: Double cœur et preuve formelle pour automatismes sil4. In:
  Congrès Lambda Mu 20 de Maîtrise des Risques et de Sûreté de
  Fonctionnement, {IJCAR} 2016, Saint Malo, France, Octobre 11 - 13, 2016,
  Proceedings (2016)

\bibitem{WCET2019}
Varoumas, S., Crolard, T.: Wcet of ocaml bytecode on microcontrollers: An
  automated method and its formalisation. In: Proceedings of the 19th
  International Workshop on Worst-Case Execution Time Analysis ({WCET} 2019).
  Stuttgart, Germany (Jul 2019)

\bibitem{ERTS2016}
Varoumas, S., Vaugon, B., Chailloux, E.: {Concurrent Programming of
  Microcontrollers, a Virtual Machine Approach}. In: {8th European Congress on
  Embedded Real Time Software and Systems (ERTS 2016)}. pp. 711--720. Toulouse,
  France (Jan 2016), \url{https://hal.archives-ouvertes.fr/hal-01292266}

\bibitem{ERTS2018}
Varoumas, S., Vaugon, B., Chailloux, E.: {A Generic Virtual Machine Approach
  for Programming Microcontrollers: the OMicroB Project}. In: { 9th European
  Congress on Embedded Real Time Software and Systems (ERTS 2018)}. Toulouse,
  France (Jan 2018), \url{https://hal.sorbonne-universite.fr/hal-01705825}

\bibitem{PADL2015}
Vaugon, B., Wang, P., Chailloux, E.: {Programming Microcontrollers in Ocaml:
  the OCaPIC Project}. In: {International Symposium on Practical Aspects of
  Declarative Languages (PADL 2015)}. Lecture Notes in Computer Science, vol.
  9131, pp. 132--148. {Springer Verlag}, Portland, OR, United States (Jun
  2015), \url{https://hal.archives-ouvertes.fr/hal-01213289}

\end{thebibliography}

\end{document}